\documentclass[runningheads]{llncs}
 \pdfoutput=1
\usepackage{graphicx}
\usepackage{proof}
\usepackage{amsmath,amssymb,textcomp}
\usepackage{dsfont}
\usepackage{tikz}
\usepackage{subcaption}
\usepackage{algorithm}
\usepackage{algorithmic}
\usepackage{cite}

\usepackage{todonotes}
\newif\ifwithnotes
\withnotestrue

\begin{document}

\title{Static extraction of memory access profiles for multi-core interference analysis of real-time tasks}

\author{Thomas Carle\inst{1} \and Hugues Cass\'e\inst{1}}

\authorrunning{T. Carle and H. Cass\'e}
\titlerunning{Memory access profiles for multi-core interference analysis}
\institute{Université Toulouse 3 Paul Sabatier, IRIT, CNRS
\email{name.surname@irit.fr}}
\maketitle        
\begin{abstract}

We present a static analysis framework for real-time task systems running on multi-core processors. Our method analyzes tasks in isolation at the binary level and generates worst-case timing and memory access profiles. These profiles can then be combined to perform an interference analysis at the task system level, as part of a multi-core Worst-Case Response Time (WCRT) analysis. In this paper we introduce a formal description of the models and algorithmic building blocks composing our framework. We also discuss how the memory access profiles generated by our method could be used to feed existing state-of-the-art WCRT frameworks. To the best of our knowledge, it is the first time that a method is documented on how to produce sound, safe and precise inputs for interference analysis methods.

    \keywords{Multicore architectures  \and Worst-Case Execution Time \and Static analysis}
\end{abstract}

\section{Introduction}
Worst-Case Execution Time (WCET) and Response Time (WCRT) analysis methods have existed for decades and are currently being used in the industry to provide static guarantees that tasks running in real-time systems will respect their deadlines. Such methods have been improved over the years in order to integrate the effects of complex hardware~\cite{otawa, absint} (e.g. pipelines, caches, branch predictors) and software~\cite{crpd} (e.g. preemption, mutual exclusion) mechanisms, but have mainly targeted single-core processors. The ongoing adoption of multi-core architectures for the implementation of hard real-time systems raises new challenges for the research community. Indeed previously unseen phenomena appear in such architectures, which can have a significant impact on the execution time of the tasks that run in parallel. This so-called timing interference stems from the fact that while tasks run in parallel on separate cores, they share some hardware resources such as memories and interconnects. Classical timing analysis methods make the hypothesis that tasks run in isolation (either on completely isolated hardware or on the same core but at separate times). When tasks run in parallel, this hypothesis no longer holds, and some additional delay can be experienced if they try to access a shared resource simultaneously. As a result their actual execution time may exceed the WCET computed in isolation, thus voiding all timing guarantees.

Different methods have been developed to handle this phenomenon, such as predictable hardware components~\cite{tcrest, pret, sic}, extensions of previously existing WCRT analysis~\cite{davis2, rt-calc} and interference-free execution models enforced through careful static scheduling and synchronization~\cite{aer}. In this paper we focus on a mixed analysis/compilation framework based on the notion of Time Interest Points (TIPs), which was first introduced in~\cite{tips}. In this framework, tasks are initially analyzed in isolation in order to pinpoint the instructions which may cause or suffer from interference at runtime (the TIPs). The result of this first phase is a representation of the worst case memory access profiles of the tasks in time, under the form of timed execution traces. This information is then abstracted as sequences of segments (one sequence per task) characterized by a worst-case duration and a worst-case number of memory accesses, to be combined in a static scheduling phase in which an interference analysis is performed. Finally, synchronizations are automatically injected in the code in order to enforce the schedule/response-time computed in the second phase. The main advantages of this method are:

\begin{itemize}
    \item It is applicable to a wide range of Commercial Off The Shelf (COTS) processors. Some restrictions apply, but are not as strict as the ones imposed in predictable hardware components,
    \item It is applicable to legacy code, with minimal automatic code modifications, where existing methods based on static scheduling require heavy transformations of the source code of the tasks to make it comply with the execution model,
    \item The byproducts of the analyses of the first phase (traces and segments) can be used to feed state-of-the-art WCRT and Real-Time calculus~\cite{davis2, rt-calc} analyses rather than a static scheduling back-end, in order to allow more dynamic implementations of the system, with minimal code adjustment (time-triggered or lock-based synchronizations, or thread yielding mechanisms and scheduler configuration).
\end{itemize}



In this paper we focus on architectures where all cores have a private scratchpad memory in which their code is loaded and are equipped with private L1 data caches and a shared memory bus implementing a greedy first-come first-served policy. In this context we provide a formal description of the models and algorithms which allow the abstraction of tasks binary code into time and memory access profiles, and discuss how these profiles can be fed to state-of-the-art analysis techniques for which, to the best of our knowledge, no method was yet provided to produce inputs.

\section{Related works}\label{sec:SOA}

The real-time systems community has been working on the problem of multi-core interference for nearly two decades now. A comprehensive survey on the topic has been published in \cite{maiza:survey}. In this section we position our work within the state-of-the-art, and focus on two existing analysis frameworks for which our results can be particularly useful.

\paragraph{\textbf{Reduction of interference through predictable execution:}} The framework we present here can be seen as a generalization of the PRedictable Execution Model (PREM)~\cite{prem} for multi-core architectures, or as a relaxation of the constraints of the Acquisition-Execution-Restitution (AER)~\cite{aer, rew} execution model. The original idea of PREM was to avoid interference between memory accesses and asynchronous I/O traffic on a bus by carefully scheduling and enforcing the execution of tasks so that it does not occur in parallel with I/O interrupts or DMA transfers. The TIPs framework leverages this idea to the problem of multi-core interference analysis: the primary objective of is to generate timing and memory access profiles of real-time tasks in order to statically schedule them on multi-core processors in a way that carefully accounts for, and possibly reduces the interference between them. The AER execution model aims at suppressing all interference by construction. The idea is to separate the execution of each task into three consecutive parts: the acquisition (A) of code and data for the task, the execution (E) of the task, and the restitution (R) of the outputs of the task to the shared memory. This separation is ensured either by the programmer or by the compiler~\cite{pagetti:aerComp}. Then the tasks are statically scheduled in a way that ensures that the A and R parts of the different tasks never occur in parallel. The TIPs framework implements the same idea, but the granularity at which it works (single memory accesses) is much finer, and it does not require to compile the task as three separate parts. This has multiple advantages such as the possibility to analyse and deploy legacy code with only small, automatic modifications (for synchronizations), and the limitation of the memory overhead due to static reservation in the AER model. Another difference is that TIPs allow the construction of programs in which some amount of interference can be tolerated (and statically quantified for compositionable processors~\cite{compositionality}).

\paragraph{\textbf{WCRT analysis frameworks:}} In~\cite{davis2} the authors present a WCRT analysis framework for sporadic task systems scheduled on multi-core processors using a preemptive fixed priority algorithm (and static partitioning of tasks on the cores). The authors consider that each possible execution trace of each task in the system is available for analysis, and from this set provide precise formulas to quantify the effect of interference between tasks on the shared elements of the target processor (memories, busses, processor time). This work extends classical WCRT analyses~\cite{wcrt} by introducing new interference terms to cover the particularities of multi-core processors, and by making it possible to precisely account for the execution context of the tasks (i.e. which other tasks are running on the same core, or in parallel). These terms are computed by extracting worst-case information for any time interval of any given size on the execution traces of tasks. In~\cite{davis2} the authors discuss the empirical complexity of obtaining and manipulating the entirety of the execution traces for a task system corresponding to an industrial application. Their conclusion is that traces are a desirable abstraction of the tasks execution behavior since they can be easily manipulated and they express precisely the relation between the task and the shared resources. In particular they emphasize the fact that the worst case behavior of a task depends on its execution context, and that traces allow to exploit this. They conclude that although working on all execution traces is unfeasible for arbitrary applications, it is possible to feed the framework with a set of abstract traces which overestimate the worst case behaviors of the tasks. However nothing is said on how to obtain such an abstraction, nor on the potential costs of the various abstraction methods that could be used. 

In~\cite{rt-calc} the authors provide a method close to real-time calculus~\cite{rt-calc-original} in order to compute the WCRT of a task system on a multi-core processor. Each task is represented as a sequence of time intervals, and for each time interval, a bound on the worst case number of memory accesses performed by the task is assumed to be known. Using this information, memory access arrival curves are derived and then combined to upper-bound the interference effect in time. A method is briefly sketched to derive the time intervals, which assumes precise knowledge on the tasks behavior (in particular local best and worst case execution times), but nothing is said on how this knowledge can be acquired in practice, nor on the abstraction cost of building the time intervals this way.

In Sec.~\ref{sec:Scheduling} we discuss how the worst-case traces and the temporal segments that are generated by the TIPs framework could be good candidates to feed the analyses of~\cite{davis2} and \cite{rt-calc}. This discussion is preceded by a precise description of these models, how they can be generated, and on the various optimization objectives that can be used to tune the analysis and their potential impact on the precision of the abstracted representations of the tasks.


\section{Static analysis framework}

In this section we first provide an overview of the TIPs static analysis framework, and then focus on each of the separate transformations that compose it.
\subsection{Overview of the method and models}
The TIPs static analysis framework processes a real-time task system by a sequence of analyzes and transformations, which are detailed in the next sections:
\begin{itemize}
    \item In a first step (Sec.~\ref{subsec:TipsGraph}), each task is analyzed in isolation. Starting from the disassembled binary of a task, a Control Flow Graph (CFG) is constructed. The CFG is analyzed in order to extract TIPs, that is to say instructions which can produce or suffer from interference. In our current implementations, we focus on instructions which may generate traffic on the memory bus due to a data cache miss, but the method could be easily extended to misses from instruction caches. Other potential sources of interference such as shared L2 caches or effects from cache coherence protocols can also be modeled in the same framework, but are left for future work. 
    \item Once the TIPs have been obtained, the CFG is transformed into a TIPsGraph (Sec.~\ref{subsec:TipsGraph} as well): a simplified control flow graph where the nodes correspond to the TIPs of the task, and the edges represent the possible control flow between the TIPs, in an abstract version. Nodes are labelled with the number of memory accesses made by the corresponding TIP, and edges are labelled with the worst case execution time of any execution paths linking the source TIP and the destination TIP of the edge. This representation is TIP-centered, and simplifies the CFG while allowing the following analyses and transformations to remain conservative.
    \item The TIPsGraph is then used to enumerate execution traces using a working list algorithm (Sec.~\ref{subsec:Traces}). The enumerated traces exhibit the occurrence of the TIPs in all possible executions of the task. For each trace, the TIPs execution dates are a worst-case approximations. The enumerated traces can be used as timed memory access profiles for the tasks in WCRT analyses, but may remain too complex to be used in practice for other methods (such as static scheduling).
    \item For uses for which the enumerated traces are too complex to be exploited, the traces for each task are then transformed into a sequence of so-called "time segments" (Sec.~\ref{subsec:Segments}): each segment has a duration and a worst case number of memory accesses, and the sequence of segments represents an over-approximation of the number of memory accesses that can be performed by the task in the corresponding time windows.
    \item In the TIPs framework, the tasks of the system are then subjected to static scheduling, using their representation as sequences of segments (Sec.~\ref{sec:Scheduling}). During this step, an interference analysis is performed, which assumes that the processor architecture is time-compositionable~\cite{compositionality}, and its results are included in the schedule. Once an acceptable schedule (i.e. which respects all real-time constraints) has been found for the whole tasks system, synchronizations are automatically inserted in the binary code of the tasks to enforce the schedule.
\end{itemize}
In the remainder of this section we will provide more details and a formal representation for each of the aforementioned steps and models.

\subsection{Extracting a TIPsGraph from a CFG}\label{subsec:TipsGraph}
The analysis of each task $\tau$ in isolation starts working on the CFG $CFG_{\tau} = \left\{\mathcal{N}, \mathcal{E}\right\}$ of $\tau$, where $\mathcal{N}$ is the set of nodes called Basic Blocks (BBs) of the graph, and $\mathcal{E}$ is the set of edges $e\in\mathcal{N}\times\mathcal{N}$ which represent the control flow of the application. In this model BBs are sequences of instructions $i_0, i_1, ..., i_n \in \mathcal{I}$ with a single entry point and a single exit point. Using MUST and MAY cache analyses~\cite{cache-analysis}, TIPs are pinpointed from the rest of the instructions. As stated before, a TIP is an instruction which may create or suffer from interference. Recall that in the scope of this paper we focus on multi-core architectures in which each core has a private L1 data cache, a private scratchpad holding the code to execute and all cores share a memory bus. In this context TIPs are the memory instructions which cannot be statically determined to always result in a hit (called in short Always Hit - AH) in the L1 data cache of the core which executes them. The objective of the first step of the analysis is to build for each task $\tau$ a TIPsGraph $TG_{\tau}=\left\{\mathcal{T}, \mathcal{E}_{TG}\right\}$ where $\mathcal{T}\subseteq \mathcal{I}\times\mathbb{N}$ is the set of TIPs of the task and $\mathcal{E}_{TG}\subseteq \mathcal{T}\times\mathcal{T}\times\mathbb{N}$ is the set of edges representing the control flow between TIPs. Each TIP $t\in\mathcal{T}$ is composed of an instruction $t.i$ and of the worst case number of memory accesses that this instruction may perform when executed $t.\mu$. Each edge $e\in\mathcal{E}_{TG}$ is composed of a couple of TIPs $(e.src,e.dst)$, as well as a conservative approximation of the worst case execution time ($e.w$) of the code portions between $e.src.i$ and $e.dst.i$.
\paragraph{\textbf{Property 1:}} $\forall e\in \mathcal{E}_{TG},$ $e=(i_j,i_k, e.w),$ $\forall p\in PATHS(i_j,i_k),$ 

\begin{center}$e.w \geq WCET(p)$, \end{center}

where $PATHS(i_j,i_k)$ is the set of possible execution paths between instructions $i_j.i$ and $i_k.i$, and $WCET(p)$ is a conservative approximation of the wcet of the code portion composed of the instructions of $p$, which can be computed using a static analysis tool.

\begin{figure*}[t]
\centering
\hspace*{-20mm}\begin{subfigure}{.8\textwidth}
\centering
\includegraphics[width=\linewidth]{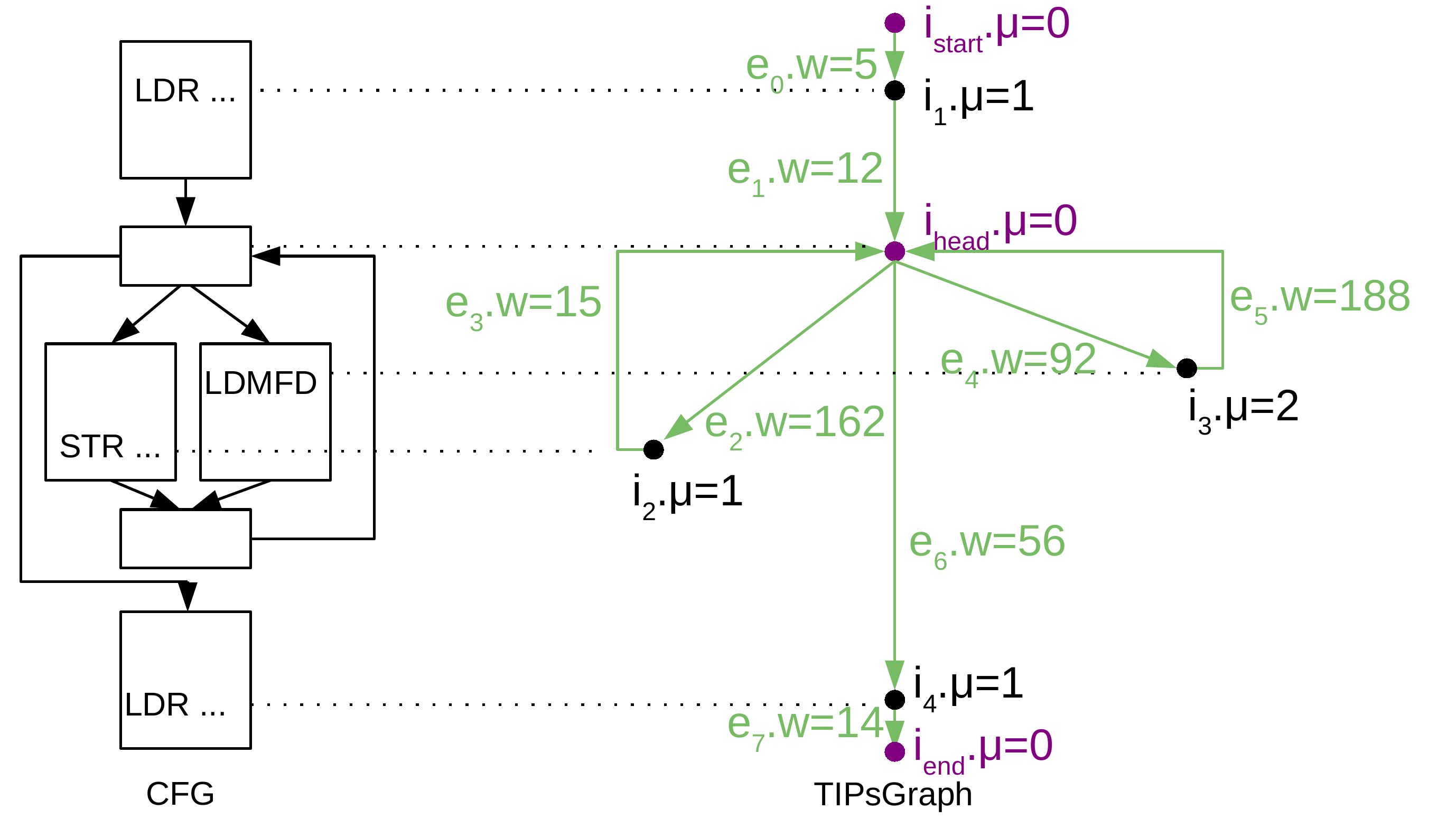}
\caption{With TIPs in the loop}\label{figure:TG1}
\end{subfigure}%
\begin{subfigure}{.8\textwidth}
\centering
\includegraphics[width=\linewidth]{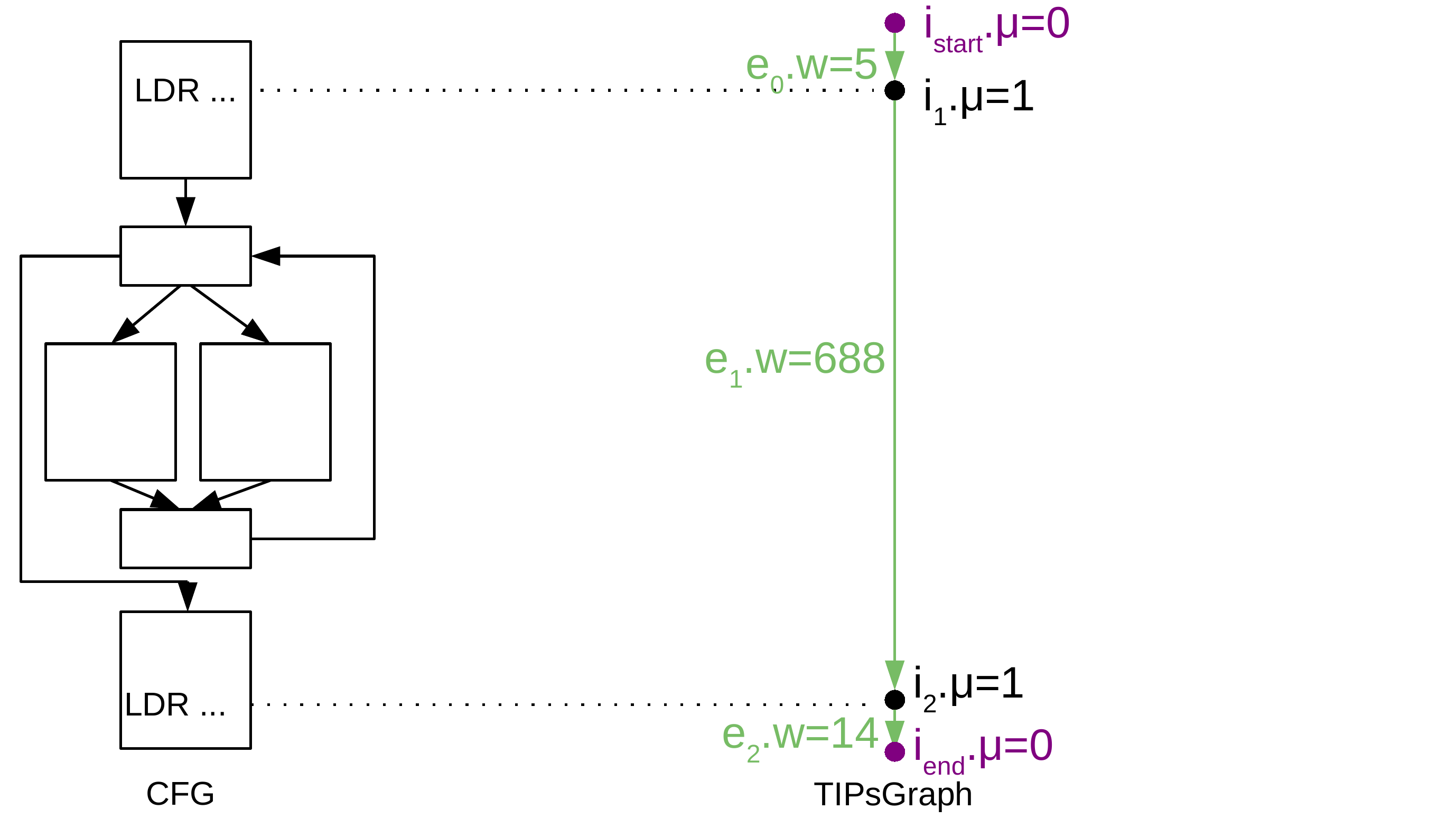}
\caption{Without TIPs in the loop}\label{figure:TG2}
\end{subfigure}
\caption{Example of CFGs and their corresponding TIPsGraphs}
\label{fig:test}
\end{figure*}

To ensure that a TIPsGraph covers the possible executions of the whole task it represents, we add two fictive nodes $i_{start}$ and $i_{end}$ which represent the entry and exit points of the task. Both $i_{start}.\mu$ and $i_{end}.\mu$ are equal to 0. Fig.~\ref{figure:TG1} shows a TIPsGraph along with the CFG from which it was extracted. The TIPsGraph starts with node $i_{start}$ and ends with node $i_{end}$. The rest of the nodes composing the TIPsGraph is extracted from the CFG: in this example we assume that four memory instructions may access the bus (the cache analysis did not result in AH for these). Each of them is represented in the TIPsGraph, as well as the possible control flow between them. Each arc records such a possible transition, and is labelled with the wcet of the portion(s) of code that are executed between the TIP instructions.

In order to correctly handle loops, a TIP $i_{head}$, with $i_{head}.\mu = 0$ is also created to represent the loop header BB, if and only if there exists at least a TIP $i$ inside the loop with $i.\mu>0$. When there is no TIP inside the loop, the loop gets abstracted in the TIPsGraph, like illustrated in Fig.~\ref{figure:TG2} : the control flow of the loop is no longer detailed in the TIPsGraph, but the edge representing the transition between the last TIP before the loop and the first TIP after the loop accounts for the worst case loop duration.

\subsection{Enumeration of timed execution traces}\label{subsec:Traces}
The next step of the analysis is to enumerate execution traces from the TIPsGraph. The result of this enumeration is an abstract representation of the possible execution traces of the task, with two interesting properties for our analysis purposes:
\begin{itemize}
    \item It is centered around memory accesses: only memory access instructions are represented (and loop headers, when the loop body contains memory access instructions) in the traces. In particular, control flow divergence which does not lead to memory accesses is abstracted away, and accounted for in the WCETs between TIPs. This reduces the empirical complexity of the subsequent analyses.
    \item All transitions between TIPs are labelled with local WCETs. This guarantees that the abstraction used to represent the tasks execution is conservative: although not all actual execution traces are detailed in the analysis, the subset on which we work is a sound conservative approximation for WCET analysis. Moreover, in combination with the following steps (static scheduling, interference analysis, injection of synchronizations), this model is also sound for the analysis of interference. 
\end{itemize}

We define a trace $tr$ as a sequence of couples $(t, d)\in \mathcal{T}\times\mathbb{N}$, where $t$ is a TIP and $d$ is a conservative approximation of the worst case execution date of $t.i$ in trace $tr$. For a trace $tr = [p_0,p_1,...,p_n]$ with $\forall i\in [|0,n|], p_i = (t_i, d_i)$, we denote by $last(tr)$ the element $p_n$. We also use $tr::p_{k+1}$ to denote the trace obtained by concatenating trace $tr$ with element $p_{k+1}$. 

\vspace*{-8mm}
\begin{figure*}
\centering
\includegraphics[width=.8\linewidth]{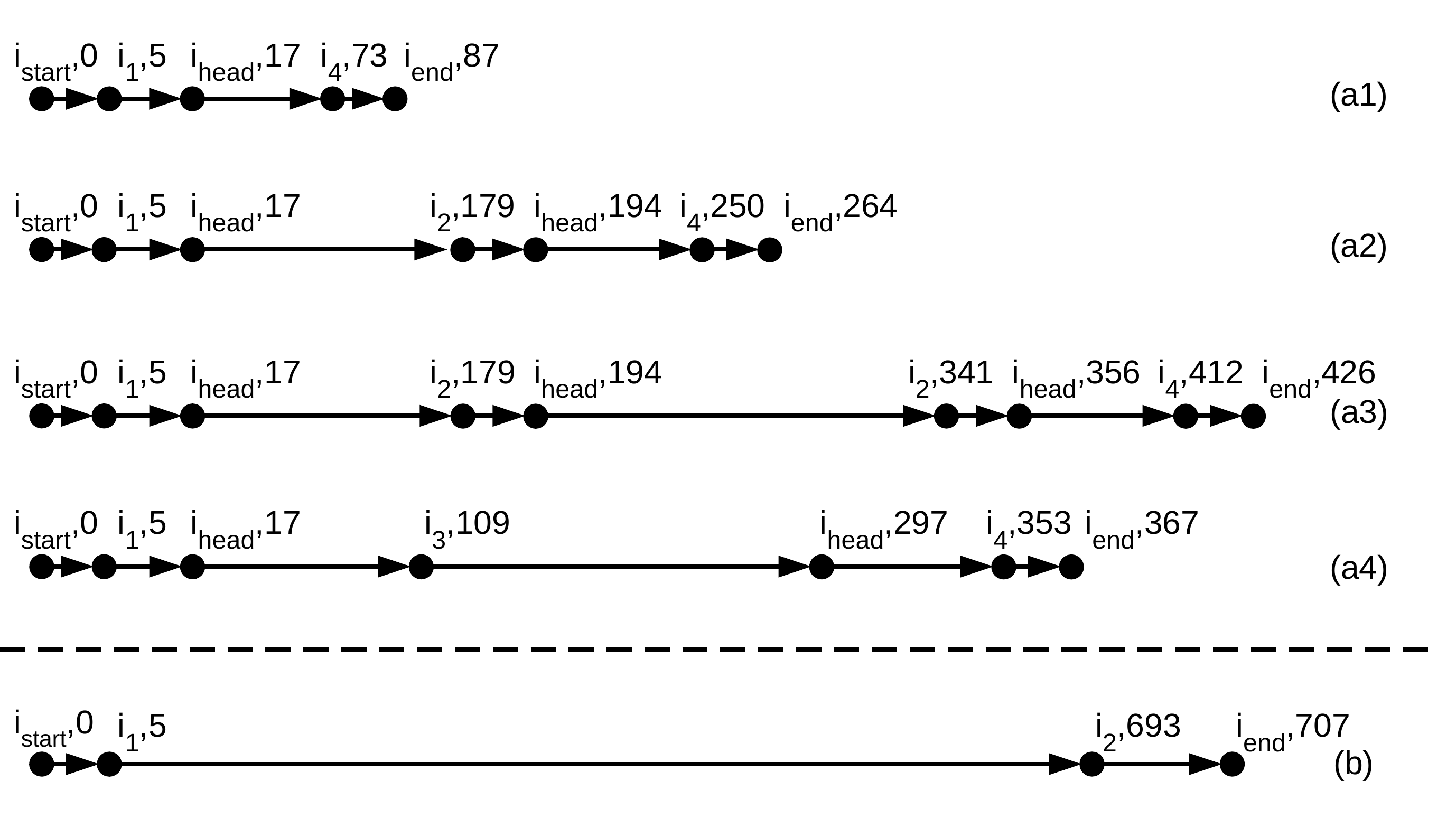}
\vspace*{-5mm}
\caption{Examples of enumerated traces from the TIPsGraphs of Fig. \ref{fig:test}}\label{figure:Traces}
\vspace*{-8mm}
\end{figure*}

\vspace{2mm}

\renewcommand{\algorithmiccomment}[1]{$\vartriangleright$ \textit{#1}}

\begin{algorithm}

\caption{Basic trace enumeration}\label{algo:WL}
\begin{algorithmic}[1]
\STATE $Traces \leftarrow \emptyset$
\STATE $tr\leftarrow (i_{start}, 0)$
\STATE $WL \leftarrow \{ (tr, e, []) ~|~ e\in\mathcal{E}_{TG} \land e.src=i_{start} \}$
\WHILE{$WL \neq []$}
\STATE $(tr, e, context)\leftarrow pop(WL)$
\STATE $(i_{last}, d)\leftarrow last(tr)$
\STATE \COMMENT{Dealing with loops}
\STATE $iteration\leftarrow pop(context)$
\IF{$is\_loop\_head(e.dst)$}
\IF{$is\_return\_arc(e)$}
\IF{$iteration = max\_bound(loop(e))$}
\STATE \textbf{continue} \hfill\COMMENT{Not a valid trace: dump it}
\ELSE
\STATE $push(context, iteration+1)$ \hfill\COMMENT{Advance iteration counter}
\ENDIF
\ELSE
\STATE $push(context,0)$ \hfill\COMMENT{Entering a new loop}
\ENDIF
\ELSE
\IF{$is\_loop\_exit(e)$}
\IF{$iteration< min\_bound(loop(e))$}
\STATE \textbf{continue} \hfill\COMMENT{Not a valid trace: dump it}
\ENDIF
\ENDIF
\ENDIF
\STATE \COMMENT{Adding a new element to the trace}
\STATE $tr\leftarrow tr::(e.dst, d+e.w)$
\IF{$e.dst=i_{end}$}
\STATE $Traces\leftarrow Traces\cup \{tr\}$
\ELSE
\STATE $WL \leftarrow WL \cup \{ (tr, e_n, context) ~|~ e_n\in\mathcal{E}_{TG} \land e_n.src=e.dst \}$
\ENDIF
\ENDWHILE
\end{algorithmic}
\end{algorithm}
A basic enumeration algorithm is described in Algo.~\ref{algo:WL}. It is a working list algorithm which performs a depth-first traversal of the TIPsGraph of a task. The working list contains triplets composed of a trace currently under construction, a TIPsGraph edge and a stack containing information regarding the current iteration of loops that are being traversed. The algorithm iteratively builds the set $Traces$ of the enumerated traces. Initially, $WL$ and $Traces$ are empty. At each step of the process, the algorithm gets a trace under construction from $WL$, along with an edge from the TIPsGraph whose source node is the current last node of the trace, and the corresponding loop iteration context. From this, the trace is extended with the destination instruction of the edge, and pushes this new state on $WL$, along with all possible successor edges of the new last node of the trace. One trace is completed and thus added to the $Traces$ set when the node $i_{end}$ has been reached.

The tricky cases concern loop headers (\textbf{L.8} to \textbf{L.26}): in order for the algorithm to finish, it is mandatory for the number of iterations of each loop of the task to be bounded (which is a basic requirement for WCET computation).
When the trace enumeration reaches an edge whose destination node corresponds to a loop header (\textbf{L.10}), the algorithm checks (\textbf{L.11}) whether the arc in question is a return arc from inside the loop (marking the end of an iteration of the loop), or not (meaning the enumeration is entering the loop for the first iteration). If the enumeration just enters the loop, a new loop iteration context is created by pushing 0 (corresponding to the first iteration of the loop) on the $context$ stack (\textbf{L.18}). The algorithm uses a stack so it can handle nested loops. If on the other hand, the current arc is a return arc, the algorithm checks if the current iteration corresponds to a valid execution : it must not exceed the maximum iteration bound for the loop. If the execution is invalid, the current trace is simply discarded (\textbf{L.13}), and the algorithm pops a new element from $WL$. In order to work, the algorithm must also be able to pop an element from the context stack when exiting a loop. This is done by detecting that the current edge exits from the loop (\textbf{L.21}), and by checking that the minimum iteration bound has been reached in the current stack (\textbf{L.22}). This minimum iteration bound is set to 0 by default, but the more precise it is, the better the outcome of the analysis.

Figure~\ref{figure:Traces} displays 5 traces enumerated from the TIPsGraphs of Fig.~\ref{fig:test}. The last trace (at the bottom), labelled (b) is the only trace that can be enumerated from the TIPsGraph of Fig.~\ref{figure:TG2}. The first element of the trace,  $i_{start}$, corresponds to the start of the execution of the task at date 0. The next elements are the execution of $i_1.i$ at date 5, the execution of $i_2.i$ at date 693 and finally the end of the task at date 707. Traces (a1) to (a4) are a subset of all possible enumerated traces from the TIPsGraph of Fig.\ref{figure:TG1}. In order to enumerate them, we assumed that the number of loop iterations varied at least between 0 iteration (trace (a1)) and 2 iterations (trace (a3)). Trace (a1) corresponds to the execution of the task when the loop is not executed. Traces (a2) and (a3) correspond to the execution of the task when the left branch of the loop is taken respectively once and twice before exiting the loop. Trace (a4) corresponds to the execution of the task when the right branch of the loop is taken once before exiting. Notice that the dates for each element of these traces are worst case dates, meaning that the corresponding instructions can in practice execute before that date, but are guaranteed to never execute after. This means that any such trace covers multiple execution patterns, which reduces the empirical complexity of the next steps of the analysis (regardless of the analysis framework). However, three important issues must be raised at this point:
\begin{itemize}
    \item Knowing only a worst case date for the memory accesses may increase the imprecision of the interference analysis, since it must consider that an access can occur at any time before the worst case date. One way to mitigate this issue is to inject synchronizations inside the code of the tasks to reduce the size of the time intervals during which accesses may occur.
    \item As illustrated by the 4 traces (a1) to (a4), the same instruction can have a different worst case date in different traces (e.g. $i_4.i$), which can also lead to imprecision in the interference analysis. Methods must be found to mitigate this issue, either at the code generation level (once again, synchronizations) or at the analysis level (careful accounting of the worst case number of memory access of the task on a given time interval). 
    \item Enumerating traces when a loop has a different minimum and maximum iteration bounds dramatically increases the empirical complexity of the enumeration algorithm: the enumeration of all possible sub-traces after the exit of the loop must be performed entirely as many times as there are ways to exit the loop (i.e. for each iteration between the minimum and maximum loop bound), even though the enumeration of these sub-traces is exactly the same each time, since they correspond to exactly the same portion of the TIPsGraph (when infeasible paths are not considered). In the given example, the portion of the TIPsGraph located after the loop is very small, but in practice we have noticed that it is not the case for arbitrary applications, and that the enumeration may become infeasible in acceptable time when the min and max bounds for a loop differ too much.
\end{itemize}


The enumerated traces are a first, rather raw representation of the timing and memory access profile of the tasks of the analyzed system. They can be used to perform a WCRT analysis following the method described in \cite{davis2}, even though caution must be taken: these are not real execution traces, but worst-case approximations of execution traces. This means that the method of \cite{davis2} will have to be adapted to take into account this specificity, or that synchronizations will have to be added to the task code in order to enforce some of the worst case dates for the memory accesses. 

In the TIPs framework however, the objective is to perform static scheduling, in order to analyze and try to limit the interference between tasks. To do so, we need to transform the enumerated traces into entities that will be practical to schedule, such as temporal segments.

\subsection{Temporal segments}\label{subsec:Segments}\vspace*{-2mm}
We now present the kind of temporal segments that are used in the TIPs framework in order to represent the time and memory access profile of tasks and to statically schedule them. A temporal segment $s_{i}$ is a triplet $(s_{i}.start,s_{i}.dur,s_{i}.\mu)$ where $s_{i}.start$ is the start date of the segment, $s_{i}.dur$ is its duration, and $s_{i}.\mu$ is a map which contains the number of memory accesses that can happen on the time interval $[s_{i}.start, s_{i}.start+s_{i}.dur]$ for each trace. In the following, we also denote by $s_{i}.end = s_{i}.start + s_{i}.dur$ the end date of segment $s_{i}$.

Any task $\tau$ (resp. any enumerated trace $tr\in Traces(\tau)$) can be abstracted using a sequence of segments $Segs_{\tau}= [s_0(\tau), ..., s_n(\tau)]$ (resp. $Segs_{tr}= [s_0(tr), ..., s_k(tr)]$), with the property that segments of a sequence do not overlap and the first segment starts at date 0 i.e. $s_0(\tau).start = 0$ and $\forall i\in[|1, n|], s_i(\tau).start\geq s_{i-1}(\tau).end$.

The shape of the segments sequence of each task will have an impact on the scheduling and interference analysis phase. A trade-off must be found between:\vspace*{-1mm}
\begin{itemize}
    \item The number of segments for each task. Scheduling elements (tasks, or segments) on a multi-core target is a NP-hard problem, so increasing the number of segments to schedule can increase the time it takes to build a schedule, potentially to a point where it is no longer feasible in practice.
    \item The length of the segments. During the interference analysis, any two segments from different tasks scheduled on overlapping time intervals on different cores will be considered as being in interference. By definition, smaller segments occupy a core for less time than larger segments, and are thus less exposed to interference from other cores. Moreover, smaller segments offer more flexibility to the scheduler to reduce the impact of interference.
    \item The worst case number of memory accesses on each segment. The length of the segments and the number and position of the synchronizations used to enforce them have an impact on the number of memory accesses attributed to each segment. This number must be conservative for each segment, so a memory access from a single instruction can be counted in multiple segments if the execution date of the instruction cannot be proven to happen in the time interval of only one segment. This can increase the imprecision of the method if one is not careful when shaping the segments and selecting the synchronization points.
    \item The number of synchronizations that will be required to guarantee that the code corresponding to the segments does not start before it is intended to. Each synchronization corresponds to additional code for the task, so their number must remain limited. Without optimization, code must be added (automatically) to the task code to ensure that in each execution trace a synchronization will be executed to enforce the start date of each segment.
\end{itemize}

In the remainder of this section, we provide algorithms that enable the extraction of valid segment representations for tasks. These are baseline algorithms which do not perform any optimization with regard to the aforementioned trade-offs. In the description of the algorithms we use the empty sequence ([]) and concatenation of an element $e$ at the right-end of a sequence $seq$ (::).


These algorithms rely on the $Intersect$ operator which is defined in Def.~\ref{def:intersect}. This operator computes the intersection of two segments : if the segments correspond to non-overlapping time intervals, the return value is empty. Otherwise, the operator returns a segment whose time interval is the intersection of the time intervals of the two input segments, and its summary of worst case number of memory accesses is the union of the summaries of worst case memory accesses of the input segments. Our algorithms also use the $Segments$ procedure described in Algo.~\ref{algo:segment}. This procedure transforms a trace $tr$ of a task $\tau$ into a sequence of segments in the following manner: for each node $n$ in the trace, it creates two segments: $s_1$ which starts at the date of the node, spans the worst case duration of the accesses of this node and has $s_1.\mu = \{tr:n.\mu\}$ (marking that on this time interval trace $tr$ makes at most $n.\mu$ accesses), and $s_2$ which starts just after and spans until the date of the next node and has $s_2.\mu=\{tr:0\}$. The procedure is also called with parameter $d_{max}$ which is the maximum of the dates of the last nodes of all traces of $\tau$ (i.e. the WCET of $\tau$ in the absence of interference). This is used to extend the last segment so that it spans until $d_{max}$.
\begin{definition}  $\forall s_{i}, s_{j} \in Segs, ~Intersect(s_{i}, s_{j}) =$
    \begin{align*}
    	\begin{cases}
    		\emptyset
    		    & if~s_{i}.start\geq s_{j}.end \lor s_{j}.start\geq s_{i}.end\\
    		(s_{i}.start,s_{i}.dur, s_{i}.\mu \cup s_{j}.\mu)
    		    & if~s_{i}.start\geq s_{j}.start  \land s_{j}.end\geq s_{i}.end\\
    		(s_{i}.start,s_{j}.end-s_{i}.start, s_{i}.\mu \cup s_{j}.\mu)
    		    & if~s_{i}.start \geq s_{j}.start\land s_{i}.end \geq s_{j}.end\\
    		(s_{j}.start,s_{i}.end-s_{j}.start, s_{i}.\mu \cup s_{j}.\mu)
    		    & if~s_{i}.start < s_{j}.start\land s_{i}.end \leq s_{j}.end\\ 
    		(s_{j}.start,s_{j}.dur,s_{i}.\mu \cup s_{j}.\mu)
    		    & if~s_{i}.start < s_{j}.start \land s_{i}.end > s_{j}.end\\
    	\end{cases}
    \end{align*}
    \label{def:intersect}
\end{definition}

\vspace*{-3mm}The top-level algorithm is described in Algo.~\ref{algo:task-seg}: starting with an arbitrary trace $tr_1$ from the set of traces of $\tau$, it transforms $tr_1$ into a segments representation using procedure $Segments$ (described in Algo.~\ref{algo:segment}): $Segs_\tau$. Then each other trace $tr_i$ of $\tau$ is transformed into a sequence of segments, and $Segs_\tau$ is updated with the intersection of the current segments of $Segs_\tau$ and the segments that represent $tr_i$. 
\begin{algorithm}
\caption{Segments creation for a task}\label{algo:task-seg}
\begin{algorithmic}[1]
\REQUIRE $ Traces(\tau)$, $tr_1\in Traces(\tau)$, $\Delta\in\mathbb{N}$, $d_{max}\in\mathbb{N}$
\ENSURE $Segs_{\tau}$
\STATE $Segs_{\tau}= Segments(tr_1, d_{max})$
\FORALL{$tr_i\in Traces(\tau)$, $tr_i\neq tr_1 $}
\STATE $Segs_{\tau}\leftarrow Intersect(Segs_{\tau},Segments(tr_i, d_{max}))$
\ENDFOR
\STATE $Segs_{\tau}\leftarrow Fusion(Segs_{\tau}, \Delta)$
\RETURN $Segs_{\tau}$
\end{algorithmic}
\end{algorithm}
When this is done, a procedure tries to reduce the number of segments using a minimum size $\Delta$, by :
\begin{itemize}
    \item preserving all segments $s$ with $max\_access(s)=0$ and $ s.dur\geq\Delta$,
    \item for all other segments, fusing consecutive segments until the result of the fusion has a length of at least $\Delta$ or there is no more available segment to fuse. When fusing segments, the information about the worst case number of memory accesses is combined trace-wise instead of blindly summed in order to limit over-approximations.
\end{itemize}

\begin{figure*}
\centering
\includegraphics[width=.7\linewidth]{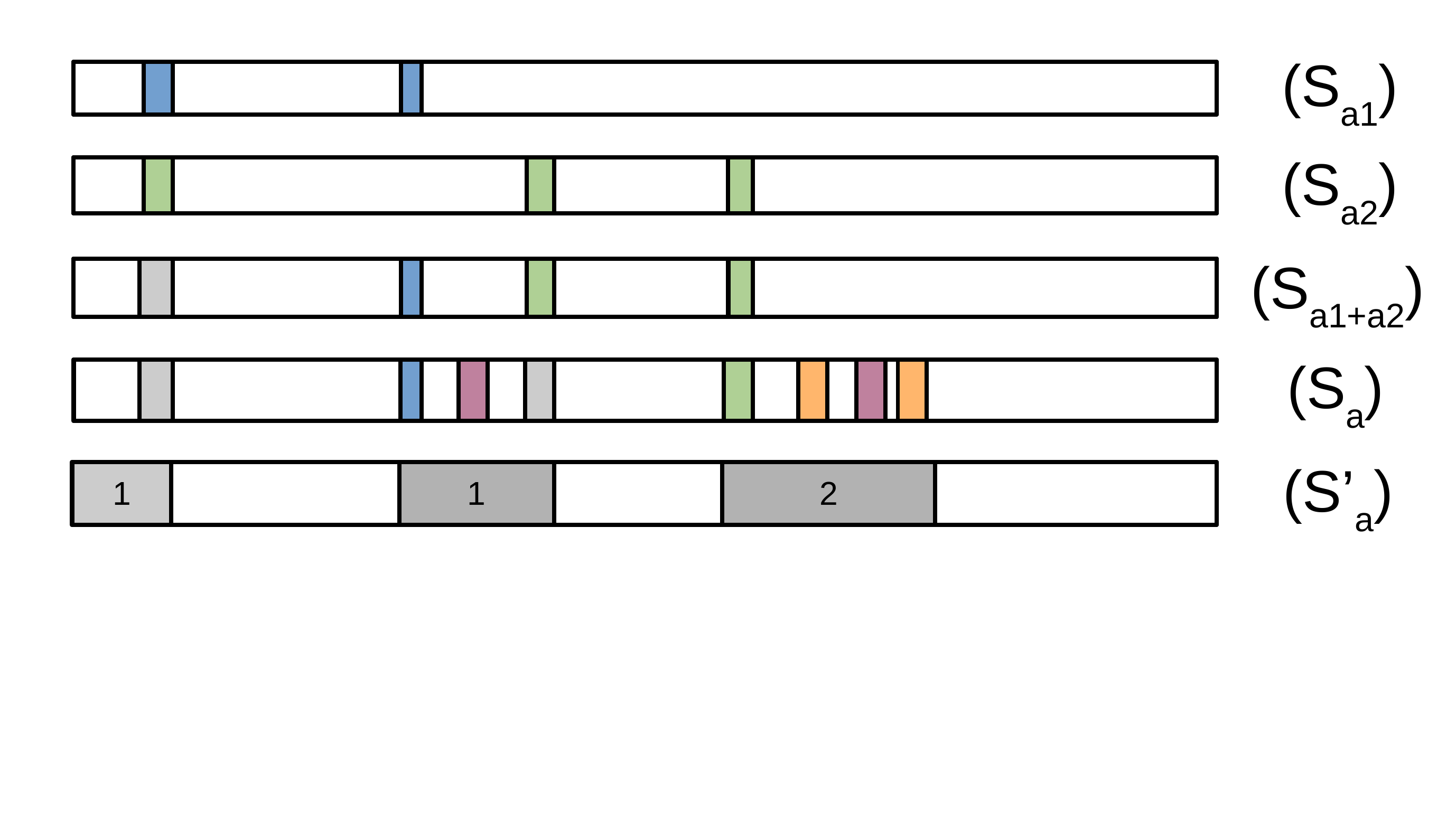}\vspace*{-15mm}
\caption{Examples of memory access profiles obtained from the traces of Fig. \ref{figure:Traces}}\label{figure:Segments}
\vspace*{-6mm}
\end{figure*}

We illustrate this algorithm in the examples of Fig.~\ref{figure:Segments}. Segment sequences ($S_{a1}$) and ($S_{a2}$) are extracted directly from traces (a1) and (a2) of Fig.~\ref{figure:Traces} using Algo.~\ref{algo:segment}. The result of their intersection is provided as ($S_{a1+a2}$). In this sequence, the first access is displayed in gray to show that this segment corresponds to either one access from trace (a1) or one access from trace (a2). The sequence labelled ($S_a$) is obtained by iterating the intersection of the traces (a1), (a2), (a3) and (a4). Different colors mean that accesses from different traces may occur. Finally trace ($S^{'}_{a}$) is obtained by fusing together the smaller segments and preserving large segments which are guaranteed to not perform any memory access. 

\begin{algorithm}
\caption{Extract a segments representation for a single trace}\label{algo:segment}
\begin{algorithmic}[1]
\REQUIRE $ tr_1\in Traces(\tau), tr_1=(i_0, d_0), (i_1, d_1), ..., (i_n, d_n)$; $d_{max}$
\ENSURE $Segs_{tr_1}$
\STATE $Segs_{tr_1}=[]$
\FORALL{$k\in[|0,n-2|] $}
\STATE $acc\_end\leftarrow d_k+i_k.\mu \times access\_time$
\STATE $Segs_{tr_1}\leftarrow Segs_{tr_1}::(d_k, acc\_end, \{tr_1:i_k.\mu\})::(acc\_end,d_{k+1}-acc\_end,\{tr_1:0\})$
\ENDFOR
\STATE $acc\_end\leftarrow d_{n-1}+i_{n-1}.\mu \times access\_time$
\STATE $Segs_{tr_1}\leftarrow Segs_{tr_1}::(d_n-1, acc\_end, \{tr_1:i_k.\mu\})$
\STATE $Segs_{tr_1}\leftarrow Segs_{tr_1}::(acc\_end, d_{max}-acc\_end, 0)$
\RETURN $Segs_{tr_1}$
\end{algorithmic}

\end{algorithm}
\vspace*{-12mm}

\vspace*{-5mm}\section{Exploitation of the memory access profiles}\label{sec:Scheduling}\vspace*{-2mm}
In this section we conclude the description of the TIPs framework by a quick discussion about the uses that can be made of the enumerated traces and segments in order to perform an interference-aware analysis of the task system. We thus discuss how the enumerated traces and segments representations can be fed to various "back-ends" such as static schedulers or the WCRT analysis frameworks described in Sec.~\ref{sec:SOA}.
\vspace*{-3mm}
\subsection{Interference analysis for static scheduling}
\vspace*{-3mm}
Regardless of the task model (single period, sporadic, multi-periodic, dependent or independent tasks) any existing static scheduling method can be adapted to the TIPs model: instead of scheduling one time interval for a task as in classical models, all segments for a task are scheduled in order. The potential interference is computed using the information contained in the segments, and must reflect the bus arbitration policy. Only segments scheduled on separate cores and whose time intervals overlap are considered to interfere. The potential interference is accounted for either by increasing the size of the segments on-the-fly (e.g.~\cite{grenoble, tips}), or by consuming an interference budget which can be added to the tasks' WCET prior to scheduling~\cite{dumitru} (e.g. a 10\% overhead on the computed WCET for interference tolerance). 
\vspace*{-5mm}
\subsection{Multi-core WCRT analysis techniques}
\vspace*{-3mm}
In order to use the interference formulas presented in~\cite{davis2} (at least for the part regarding interference on the buses and memories), an upper bound on the number of accesses that can be made in any time-interval of any size must be found for each task. In the TIPs framework, this information is available at 2 different abstraction levels: the enumerated worst-case traces (Sec.~\ref{subsec:Traces}) which offer a finer level of granularity, and thus can lead to more precise bounds at the cost of a more complex computation, and the segments (Sec.~\ref{subsec:Segments}) which represent the memory access profiles at an higher level of abstraction. In both cases, since the timings obtained in the TIPs framework are worst-case dates, synchronizations must be added to the code to be able to lower-bound with certainty the occurrence of accesses in time. The same is true for the real-time calculus method of \cite{rt-calc}, for which the segment representation obtained through the TIPs framework is a natural input format.
\vspace*{-4mm}

\vspace*{-1mm}
\section{Conclusion and future works}\vspace*{-4mm}
We presented the TIPs framework: a collection of models and algorithms for the extraction of precise timing and memory access profiles of real-time tasks. For each level of abstraction, we provided a formalization of the corresponding models and algorithms, as well as a discussion on the cost of the presented abstractions. We finally discussed how the obtained memory access profiles could be used as inputs for existing state-of-the-art analysis frameworks and tools.
In the future, we will work on optimizations and the evaluation of their impact on the different "back-ends". In particular, we are currently working on a multi-criterion optimization of the transformation of enumerated traces into segments, trying to minimize the overestimation of the number of accesses in each segment and the number of synchronizations to add to the tasks' code. 
\vspace*{-5mm}
\bibliographystyle{splncs04}
\bibliography{main}

\end{document}